\documentclass[aip,cha,reprint,numerical,onecolumn]{revtex4-1}
\usepackage{amsmath}
\usepackage{amssymb}
\usepackage{graphicx}
\usepackage{dcolumn}
\usepackage{bm}
\usepackage{natbib}		
\usepackage{color}

\newcommand {\e} {\varepsilon}
\newcommand {\vp} {\varphi}
\newcommand {\w} {\omega}
\newcommand{\dd}{\mathrm{d}}

\begin{document}

\title{Dynamical disentanglement in an analysis of oscillatory systems: an application 
to respiratory sinus arrhythmia}

\author{M. Rosenblum$^{1,2}$, M. Fr\"uhwirth$^{3}$, M.~Moser$^{3,4}$, and A. Pikovsky$^{1,2}$\\
$^{1}$Institute of Physics and Astronomy, University of Potsdam, 
Karl-Liebknecht-Str. 24/25, 14476 Potsdam-Golm, Germany\\
$^{2}$Control Theory Department, Institute of Information Technologies,
Mathematics and Mechanics, Lobachevsky University Nizhny Novgorod, Russia\\
$^{3}$Human Research Institute of Health Technology 
 and Prevention Research, Franz Pichler Street 30, A-8160 Weiz, Austria\\
$^{4}$Physiology, Otto Loewi Research Center for Vascular Biology,
 Immunology and Inflammation, Medical University of Graz,
  Neue Stiftingtalstr. 6/D05, A-8010 Graz, Austria}


\begin{abstract}
We develop a technique for the multivariate
data analysis of perturbed self-sustained oscillators. 
The approach is based on the reconstruction of the phase dynamics model from observations 
and on a subsequent exploration of this model. For the system, 
driven by several inputs, we suggest a
dynamical disentanglement procedure, allowing us to reconstruct the variability of the system's 
output that is due to a particular observed input, or, alternatively, to
reconstruct the variability which is caused
by all the inputs except for the observed one. We focus on the application of the method 
to the vagal component of the heart rate variability caused by a respiratory influence. 
We develop an algorithm
that extracts purely respiratory-related variability, using a respiratory trace and times 
of R-peaks in the electrocardiogram. The algorithm can be applied to other systems 
where the observed bivariate data can be represented 
as a point process and a slow continuous
signal, e.g. for the analysis of neuronal spiking.
\end{abstract}
\maketitle

\section{Introduction}
One of the basic problems in the data analysis is to select or to eliminate 
a particular component of a given time 
series, e.g. to remove noise or a trend, or to single out an oscillation in  
a certain frequency band, etc. A whole 
variety of techniques has been designed 
to tackle this task by means of filtering in the frequency domain, smoothing 
in a running window, subtracting a fitted polynomial, and so on. 
Furthermore, a number of modern methods -- principal component analysis, 
independent mode decomposition, empirical mode decomposition 
\cite{Fukunaga-90,Huang-98,Jolliffe-2002,Flandrin-04,Feldman-11,PhysRevE.92.032916}
-- represent 
a signal of interest as a sum of modes such that (at least) dominating modes are 
assumed to represent certain relevant dynamical processes. Correspondingly, some of these 
modes can be analyzed separately or, on the contrary, if they are considered
as irrelevant, they can be subtracted from the original data, so that 
the cleansed signal can be further processed. 

\begin{figure}[!h]
\centering\includegraphics[width=\textwidth]{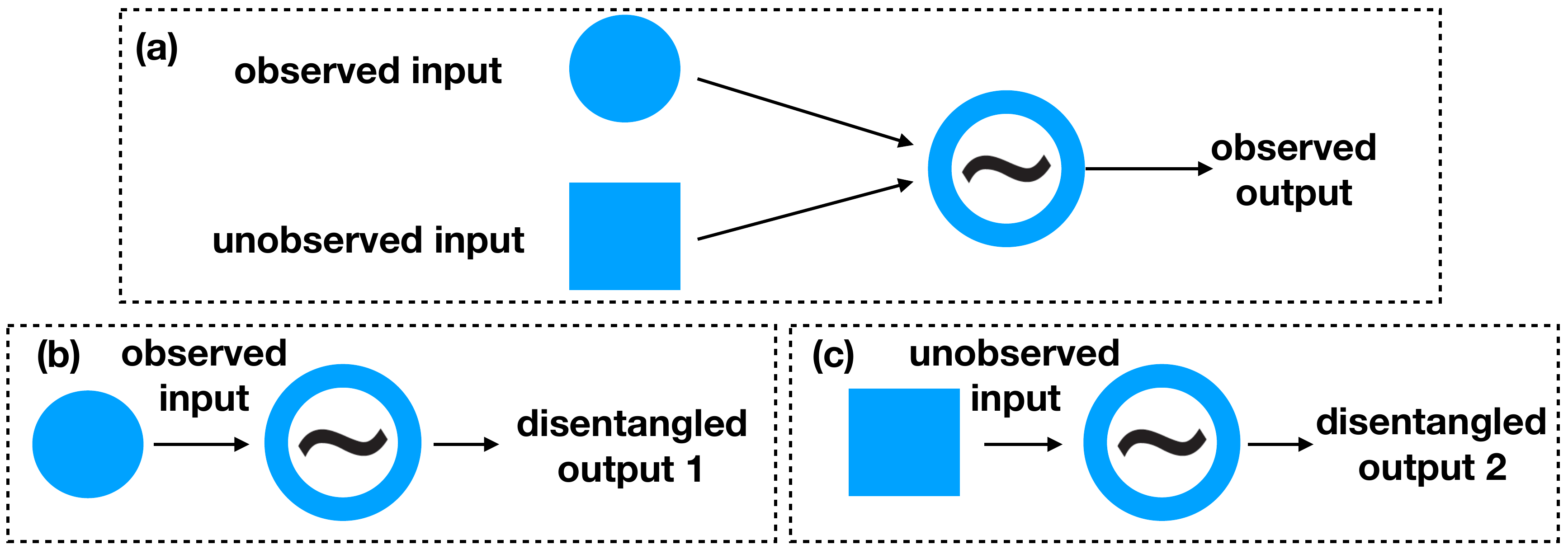}
\caption{Schematic representation of the disentanglement procedure. (a): Original
setup, where two inputs influence the oscillator. (b): Reconstruction of the
``noise-free'' dynamics. (c): reconstruction of the ``signal-free'' dynamics.
}
\label{fig:sch}
\end{figure}

In this publication we elaborate on a technique for a \textit{dynamical
disentanglement of
different components}, designed for the analysis of signals,
generated by coupled oscillatory systems. The disentanglement task
is illustrated in Fig.~\ref{fig:sch}. We assume that a signal from 
an oscillatory unit, which is driven by an observed nearly periodic signal
and by other, non-observed inputs  is known (Fig.~\ref{fig:sch}a). 
(We treat the unobserved input as some noise, although
generally it may contain some regular components as well.) 
The technique is based on a reconstruction 
of the phase dynamics of the analyzed unit. The obtained equation is then used
for generation of two new outputs. 
If only the observed input is used, i.e. the unobserved noise 
term is omitted, then the simulated equation yields a signal 
representing the dynamics of the noise-free
system, i.e. the system driven by the observed input only (Fig.~\ref{fig:sch}b). 
If, on the contrary, we eliminate from the equation the observed input, then 
the simulations yield the noise-induced output (Fig.~\ref{fig:sch}c).
This disentanglement procedure is neither the standard filtering (because 
the preserved and eliminated components can overlap in the frequency domain), nor
the mode decomposition (because the sum of two disentangled outputs 
does not yield the original signal). 
Here we consider application of this approach to cardiac and respiratory data 
in humans. Our main oscillatory unit will be the cardiovascular system, 
and the observed input will be respiration. 
As the results of the analysis we will obtain two heart rate variability
signals: one influenced purely by respiration, and one where the influence of
respiration is excluded.

Understanding of the cardiac dynamics in terms of coupled oscillators
goes back to the pioneering work by van der Pol and van der 
Mark~\cite{van_der_Pol_van_der_Mark-28}.
Within the last two decades this idea was widely used to address the interaction 
between the cardiovascular and respiratory systems with the aim 
to reveal and quantify synchronization between them and to infer
directionality and strength of their coupling 
~\cite{Schaefer-Rosenblum-Kurths-Abel-98,Mrowka-Patzak-Rosenblum-00,%
PhysRevLett.85.4831,Rosenblum_et_al-02,Mrowka_et_al-03,Kralemann_et_al-13,Iatsenko_et_al-13}.
Here we discuss how application of the coupled oscillators theory 
helps in the analysis of the main effect of the cardio-respiratory interaction,
namely modulation of the heart rate by respiration, known for about a century 
as respiratory sinus arrhythmia 
(RSA)~\cite{Eckberg-83,Berntson-Cacioppo-Quigley-93,Eckberg-03,Billman-11,Beauchaine-15}.

The separation and proper quantification of this respiratory component of 
heart rate variability (HRV) is 
of great importance for both fundamental physiological research and clinical 
medicine~\cite{Lehofer1999,Moser1998}, due to the role vagal activity 
plays not only in cardiovascular, but also in inflammatory control~\cite{Tracey2002}. 
The isolated immune system is over-reactive and self propagating by it's nature. 
Germs or degraded cells in our body are detected by immune cells like macrophages 
floating in the interstitial space of the tissue. 
Macrophages detecting germ intruders produce inflammatory signals such as TNF-alpha and 
interleucine 1~\cite{Olofsson_et_al-12}, 
which attract other immune cells from nearby blood vessels. 
Without neuro-humoral control, the immune system would enter a dangerous state of 
generalized inflammation, well known as ``sepsis'' in clinical medicine. 

To prevent this generalized reaction, vagal afferents (transmitting from periphery to brain) 
also carry receptors for these signal substances and transmit the information on inflammation 
location and strength to brain stem areas~\cite{Andersson-Tracey-12}. 
After processing this information, vagal efferents (transmitting from brain to periphery) respond 
by release of acetylcholine at the location of the inflamed tissue~\cite{Tracey2002}. 
Nicotinergic acetylcholine receptors have been identified on the surface of the macrophages, 
which down-regulate the cytocine production as a response to the cholinergic 
stimulation~\cite{Rosas-Ballina-Tracey-09}, 
thereby reducing the attraction of additional inflammatory immune cells and down-regulating 
immune response. This inflammatory feedback loop prevents over-activity of the immune system 
enabling the brain to locally control the immune activity. Therefore, a reduction of the vagal tone, 
e.g., by different forms of stress, is suspected to be related to several chronic diseases 
induced by inflammation, including type 2 diabetes, ulceral colitis, Hashimoto's thyreoiditis, 
and even cancer~\cite{Nathan-Ding-2010}. 
Severe reductions of vagal tone has been observed in patients with these 
conditions~\cite{Donchin_et_al-92,Moser2006,Das2011,Chow_et_al-14}. 
The action of sympathetic activity in this system is not as well-understood as vagal contribution 
at the moment. Therefore it is important to measure the vagal component separated from the other
components.  Linear separation by filtering the signal can improve the estimation of pure vagal tone, 
but under certain conditions may fail to do so, when the respiratory frequency approaches other 
meta-cardiac cycles deriving from sympathetic origin, like the blood pressure rhythm of 0.1 Hz. 

In our previous publications \cite{Kralemann_et_al-13,Topcu_et_al-18}
we applied the dynamical disentanglement approach to the analysis of RSA in heart rate variability 
records. In these publications we used simultaneous measurements of 
electrocardiogram (ECG) and 
respiratory activity in order to 
reconstruct the equation of the phase dynamics of the cardiac oscillator. Next, we 
exploited this equation for a decomposition of the heartbeat 
intervals series into respiratory-related and non-respiratory-related components.
This decomposition can be used as a general preprocessing tool for quantification of
respiratory related heart rate variability and, in particular, opens a new way 
to address the clinically important problem of RSA quantification.

However, the results of Refs.~\cite{Kralemann_et_al-13,Topcu_et_al-18} can be considered 
only as a proof of principle, because they were obtained using a 
\textit{continuous phase} of 
the cardiac oscillators. Determination of such a phase requires very clean high-quality
measurements and a tedious preprocessing.  Here we suggest an easy-to-implement practical 
algorithm for achieving the same goal using the information about timing of the
R-peaks only. The latter are well-defined events within each cycle of cardiac activity 
and they can be readily obtained with any standard equipment. From the viewpoint of data 
analysis, we deal with a relatively slow smooth signal (respiration), the phase of which
can 
be easily estimated, e.g. by means of the Hilbert transform, and a point process 
(R-peaks) with a frequency about 3 times higher. 
Point processes frequently appear in neuronscience, and, thus, our algorithm 
can be also helpful in the analysis of neural data, e.g. of spiking 
neurons affected by a slow observed continuous force.

\section{Dynamical disentanglement based on the phase dynamics modeling}
Our general goal is to identify dynamical properties of an oscillatory system,
related to different influences, from the
observations of its behavior in a complex noisy environment.
For example, one can be interested in the following questions:
what would be the dynamics of the system if it were noise-free?
Or, how the statistical properties of the oscillation would change if 
one of the external forces were switched off?
We address these and similar problems using the phase dynamics 
theory, see, e.g.~\cite{Winfree-67,Kuramoto-84,Pikovsky-Rosenblum-Kurths-01}.

Consider a limit cycle oscillator, weakly perturbed by regular or
stochastic \textit{known} forces $\eta_k(t)$, $k=1,2,\ldots$.
Then, according to the theory,
in the first approximation in amplitude of these forces, 
the phase dynamics obeys 
\begin{equation}
\dot\vp=\w+\sum_k Q_k[\vp,\eta_k(t)]+\zeta(t)\;.
\label{eq:gpd}
\end{equation}
Here $\vp(t)$ and $\w$ are the phase and the natural frequency of the system, 
and $Q_k$ are the coupling functions; they quantify response of the oscillator 
to the corresponding perturbations. The random term $\zeta(t)$ accounts for 
intrinsic fluctuations of the system parameters. 
Notice that the same equation describes dynamics of weakly chaotic systems;
in this case $\zeta(t)$ reflects effects of chaotic amplitude variations.
In the second-order approximation in the force amplitudes, one expects 
appearance of triple terms like $Q_{12}[\vp,\eta_1,\eta_2]$, 
etc~\cite{Kralemann-Pikovsky-Rosenblum-11,Kralemann-Pikovsky-Rosenblum-14},
but these effects will be neglected below.

Let us suppose first that Eq.~(\ref{eq:gpd}) is known. Then, if we are 
interested in
properties of the purely deterministic phase dynamics, we can solve 
numerically Eq.~(\ref{eq:gpd}) 
\textit{without the noise term} $\zeta(t)$ (we speak on the deterministic
dynamics here because the forces $\eta_k(t)$ are known (recordered)
functions of time, though they must not be completely regular). 
If the task is to analyze the response
of the oscillator to a particular external force, e.g. $\eta_1(t)$, then we 
omit in Eq.~(\ref{eq:gpd}) the terms $\xi(t)=\sum_{k>1} Q_k[\vp,\eta_k(t)]+\zeta(t)$, 
simulate the equation
\begin{equation}
\dot\vp=\w+Q_1[\vp,\eta_1(t)]\;,
\label{eq:gpd1}
\end{equation}
and analyze the obtained result according to a particular problem in question.
This approach was used in \cite{Rosenblum-Pikovsky-18} for reconstructing  the
Arnold tongue of a noise-free oscillator (with strictly
regular force $\eta_1(t)$) from a measurement of noisy  system (where
in addition to $\eta_1(t)$ also pure noise $\zeta(t)$ is present).
Alternatively, if we are interested in the effects of the random 
component $\zeta(t)$, e.g. in properties of phase diffusion, then we have to 
omit the deterministic perturbations and solve numerically 
\begin{equation}
\dot\vp=\w+\xi(t)\;.
\label{eq:gpd1a}
\end{equation}
In this way we achieve the desired dynamical disentanglement. 
Below we apply this general idea to the analysis of cardio-respiratory interaction.

\section{Disentanglement of the heart rate variability}	
In Ref.~\cite{Kralemann_et_al-13} we used the 
measurements of ECG and respiratory flow from healthy adults
in order to reconstruct the model of cardiac phase dynamics in the form 
\begin{equation}
\dot\vp=\w+Q(\vp,\psi) +\xi(t)\;,
\label{Qcard}
\end{equation}
where $\vp$ and $\psi$ correspond to the instantaneous 
phases of the cardiac and the respiratory rhythms, respectively.
This equation is a particular case of Eq.~(\ref{eq:gpd}),
with $\eta_1(t)$ corresponding to the respiration dynamics. Since the latter is 
a rhythmical process with a well-defined phase $\psi$, we write the corresponding 
coupling function as a function of two phases, $Q_1(\vp,\eta_1(t))=Q(\vp,\psi)$,
while the contribution of other, unobserved, perturbations and of intrinsic 
fluctuations is combined in the rest term 
$\xi=\sum_{k>1}Q_k(\vp,\eta_k(t))+\zeta(t)$. 
Practically, $Q(\vp,\psi)$ as a function
of two variables was constructed on a $64\times 64$ equidistant grid  on a domain
$(0,2\pi)\times(0,2\pi)$.

Notice that determination of the respiratory phase $\psi$ is simple: 
since the respiratory signal looks like a modulated and slightly distorted 
sine-wave, its phase can be easily 
estimated, e.g., by means of the Hilbert transform. 
On the contrary, the ECG signal has a 
quite complicated form and computations of its phase represent 
a nontrivial stand-alone problem, see  Ref.~\cite{Kralemann_et_al-13}:
here one needs very 
high-quality data, and its processing is technically quite demanding. 
This fact motivates a development of techniques operating only with point 
processes, namely with instants of the R-peaks, corresponding to the peak of 
depolarization of the ventricles of the human heart. 
These events can be easily detected and therefore 
are commonly used in the analysis of HRV. 
Since these peaks appear once per heartbeat cycle, their continuous phase $\vp$
without loss of generality
can be set to zero.

First, we discuss how the disentanglement of the HRV can be performed
if both continuous phases $\vp(t)$ and $\psi(t)$ are available~\cite{Kralemann_et_al-13}. 
For this goal we notice that, 
for a  given time series $\vp(t)$ and $\psi(t)$,
the coupling function in Eq.~(\ref{Qcard}) can be also interpreted
as a time series $Q[\vp(t),\psi(t)]=Q(t)$.
Correspondingly, knowledge of time series $\dot\vp(t)$ and $Q(t)$ 
yields the rest term $\xi(t)=\dot\vp-Q$. Having all these time series, we easily 
construct the new disentangled ones. These are the respiratory-related (R) and the
non-respiratory 
related (NR) components of the instantaneous cardiac frequency, denoted
as $\dot\vp^{(R)}$ and $\dot\vp^{(NR)}$, and obtained according to equations
\begin{equation}
\dot\vp^{(R)}=\w+Q(\vp^{(R)},\psi)\quad\text{and}\quad \dot\vp^{(NR)}=\w+\xi(t)\;.
\label{eq_dishrv}
\end{equation}
Notice that this is not a simple decomposition because 
$\dot\vp(t)\ne \dot\vp^{(R)}(t)+\dot\vp^{(NR)}(t)$.
In Ref.~\cite{Kralemann_et_al-13} we have demonstrated that power spectrum 
of $\dot\vp^{(R)}$ nicely describes the spectral peaks corresponding to the 
frequency of respiration and to the side-bands of the heart rate. 

In the subsequent study~\cite{Topcu_et_al-18}, we extended this idea and 
generated artificial sequences of heartbeat events (R-peaks) 
according to the conditions $\vp^{(R)}(t_k^{(R)})=2\pi k$ 
and $\vp^{(NR)}(t_k^{(NR)})=2\pi k$, $k=1,2,\ldots$,
where the phases were obtained via numerical integration\footnote{For integration 
we used the Euler scheme; 
for initial conditions both $\vp^{(R)}$
and $\vp^{(NR)}$ we set to zero at the instant of the first R-peak in the original 
data set. Since the coupling function $Q$ is given on a grid, spline interpolation 
was used to compute $Q(\vp,\psi)$ for arbitrary $\vp,\psi$.} of differential 
Eqs.~(\ref{eq_dishrv}).
The point process $t_k^{(R)}$ represents instants of the heart beats as they would 
appear if there were no other perturbations to the cardiac oscillator, except for the 
respiration, while  $t_k^{(NR)}$ represents the heart rate variability due to internal 
fluctuations and external non-respiratory rhythms, e.g. blood pressure and 
blood perfusion rhythms. 
It has been suggested that described decomposition into respiratory-related (R-HRV)
and non-respiratory-related (NR-HRV) components shall be used as a generic 
preprocessing technique prior to a quantification of the RSA in clinical practice.
This suggestion has been supported by computation of different measures of RSA
from the original series of inter-beat intervals as well as from 
respiratory-related intervals $T^{(R)}=t_{k+1}^{(R)}-t_k^{(R)}$, 
see~\cite{Topcu_et_al-18} for details.
Notice that our approach is intrinsically nonlinear, in contrast to 
\textit{ad hoc} techniques used for the same purpose, 
like adaptive filtering and least-mean-square fitting of power 
spectra~\cite{Widjaja_et_al-14,Kuo-Kuo-16}.

Summarizing, the disentanglement of the instantaneous cardiac frequency 
into R-HRV and NR-HRV components can be easily implemented, provided the continuous
phases $\vp,\psi$ are known. 
However, as already mentioned, computation of the instantaneous cardiac phase
$\vp$ requires high-quality measurements,
visual inspection of the data, extensive preprocessing,  and is currently solved 
by \textit{ad hoc}, not automated, techniques only.  
On the other hand, determination of the R-peaks
is a standard task and can be easily accomplished. Therefore, development of an
disentanglement algorithm for the case when one observable, e.g. respiration, is
continuous and appropriate for the phase estimation, and the other one, e.g. heartbeats, 
is a point process, represents an important unsolved problem. 
Below we present an approximate 
solution of this problem.

\section{Dynamical disentanglement for the point process data}
Our starting point is the description of the cardio-respiratory phase dynamics  
in form of Eq.~(\ref{Qcard}).
We assume that the respiratory phase $\psi$ is obtained from the respiratory
time series and that the instants $t_k$, when the R-peaks appear in the electrocardiogram, 
are determined. The cardiac phase at these instants is $\vp(t_k)=2\pi k$.
Let the inter-beat intervals be denoted as $T_k=t_{k+1}-t_k$. 
Then, assuming weakness of the coupling, 
$\parallel Q \parallel \ll \w$, where $\parallel\cdot\parallel$
denotes the norm of the function, and keeping in Eq.~(\ref{Qcard}) only 
the deterministic term, we write in the first approximation
\begin{equation}
\int_{t_k}^{t_{k+1}}\dd t=T_k=\int_0^{2\pi}\frac{d\vp}{\w+Q(\vp,\psi)}
\approx \frac{2\pi}{\w}-\frac{1}{\w^2}\int_0^{2\pi} Q(\vp,\psi)\dd\varphi\;.
\label{eq:cr}
\end{equation}
Next, since the respiration is much slower than the heart rate, we assume that 
within the inter-beat interval $T_k$, the phase $\psi$ grows linearly in time with the 
frequency  $\w^{(r)}_k$, i.e. $\psi(t)=\psi_k+\w^{(r)}_k(t-t_k)$, 
where $\psi_k=\psi(t_k)$.
Then the integral in Eq.~(\ref{eq:cr}) can be approximated as 
\[
-\w^{-2}\int_0^{2\pi} Q(\vp,\psi)\dd\vp=-\w^{-2}\int_0^{T_k} Q[\vp(t),\psi(t)]\dd t
\approx F(\psi_k,\w^{(r)}_k)\;.
\]
Taking for simplicity $\w^{(r)}_k=\dot\psi(t_k)=\dot\psi_k$ (corrections
to this expresion, due to slowness of the respiratory phase, appear in the higher orders)
and denoting $T=2\pi/\w$, we obtain
 \begin{equation}
 T_k = T+F(\psi_k,\dot\psi_k)+\chi_k\;,
\label{eq:cmap}
\end{equation}
where $F(\psi_k,\dot\psi_k)$ can be understood as a discrete version of the coupling function
(we denote it as the coupling map) and the rest term $\chi_k$ is the random component. 
Equation~\eqref{eq:cmap}
can be considered as a direct discrete analogue of continuous Eq.~\eqref{Qcard}.

Introducing the mean respiratory frequency 
$\bar\w=\langle \w^{(r)}_k \rangle_k=\langle\dot\psi_k\rangle_k$ 
and expressing $F(\psi_k,\w_k)$ as a Taylor-Fourier series, we finally write 
\begin{equation}
T_k\approx T+\sum_{n=1}^{N_F}\left\lbrace 
\left[\sum_{m=0}^{N_T-1} a_{n,m}(\dot\psi_k-\bar\w)^m\right ] \cos(n\psi_k)+
\left[\sum_{m=0}^{N_T-1} b_{n,m}(\dot\psi_k-\bar\w)^m\right ] \sin(n\psi_k)
\right\rbrace\;.
\label{eq:fit}
\end{equation}
Here $N_F$ and $N_T$ are the orders of the Fourier and Taylor series, 
respectively.
For a sufficiently long series of inter-beat intervals $T_k$,
 Eqs.~(\ref{eq:fit}) can be considered as an overdetermined linear system 
for unknown parameters $T,a_{n,m}$, $b_{n,m}$. This system can be easily 
solved, e.g., by mean squares minimization.

Thus, the suggested algorithm yields a discrete dynamical
model~\eqref{eq:cmap} for the inter-beat intervals. 
Now this model can be used for the dynamical disentanglement. 
In order to construct the respiratory-related component
we first take $t_1^{(R)}=t_1$.
Then, substituting $\psi_1$, $\dot\psi_1$ in (\ref{eq:fit}) we 
obtain $T_1^{(R)}$ and $t_2^{(R)}= t_1^{(R)}+T_1^{(R)}$. 
Next, we compute
$\psi(t_2^{(R)})$, $\dot\psi(t_2^{(R)})$ and 
use the model (\ref{eq:fit}) to obtain $T_2^{(R)}$ and $t_3^{(R)}$, and so 
on~\footnote{For a high-resolution measurement phase 
and frequency of respiration are given 
as a time series with a small time step. Therefore, their values
at  $t_2^{(R)}$ can be obtained, e.g. by linear interpolation between 
two closest data points.}.
For the construction of the non-respiratory-related component
we also start by assigning $t_1^{(NR)}=t_1$ and then proceed 
as follows. Let already determined $t_l^{(NR)}$
fulfill $t_k<t_l^{(NR)}<t_{k+1}$. For $t_k$ and $t_{k+1}$ we compute 
the rest term of the model (\ref{eq:fit}) (effective noise), i.e. the 
difference between the true $T_k$, $T_{k+1}$ and their value predicted by Eq.~(\ref{eq:fit}); 
these terms are $\chi_k$ and $\chi_{k+1}$. Then, using 
linear interpolation to find the effective noise at $t_l^{(NR)}$, we obtain
\begin{equation}
  t_{l+1}^{(NR)}=t_l^{(NR)}+T+\chi_k+
  \frac{\chi_{k+1}-\chi_k}{t_{k+1}-t_k}(t_l^{(NR)}-t_k)\;.
\end{equation}
The R-HRV component can be further used for an improved quantification of the 
RSA, while the NR-HRV time series can be exploited for the analysis of the other 
sources of the heart rate variability.

\section{Testing the approach on model data}
\begin{figure}[!h]
\centering\includegraphics[width=2.5in]{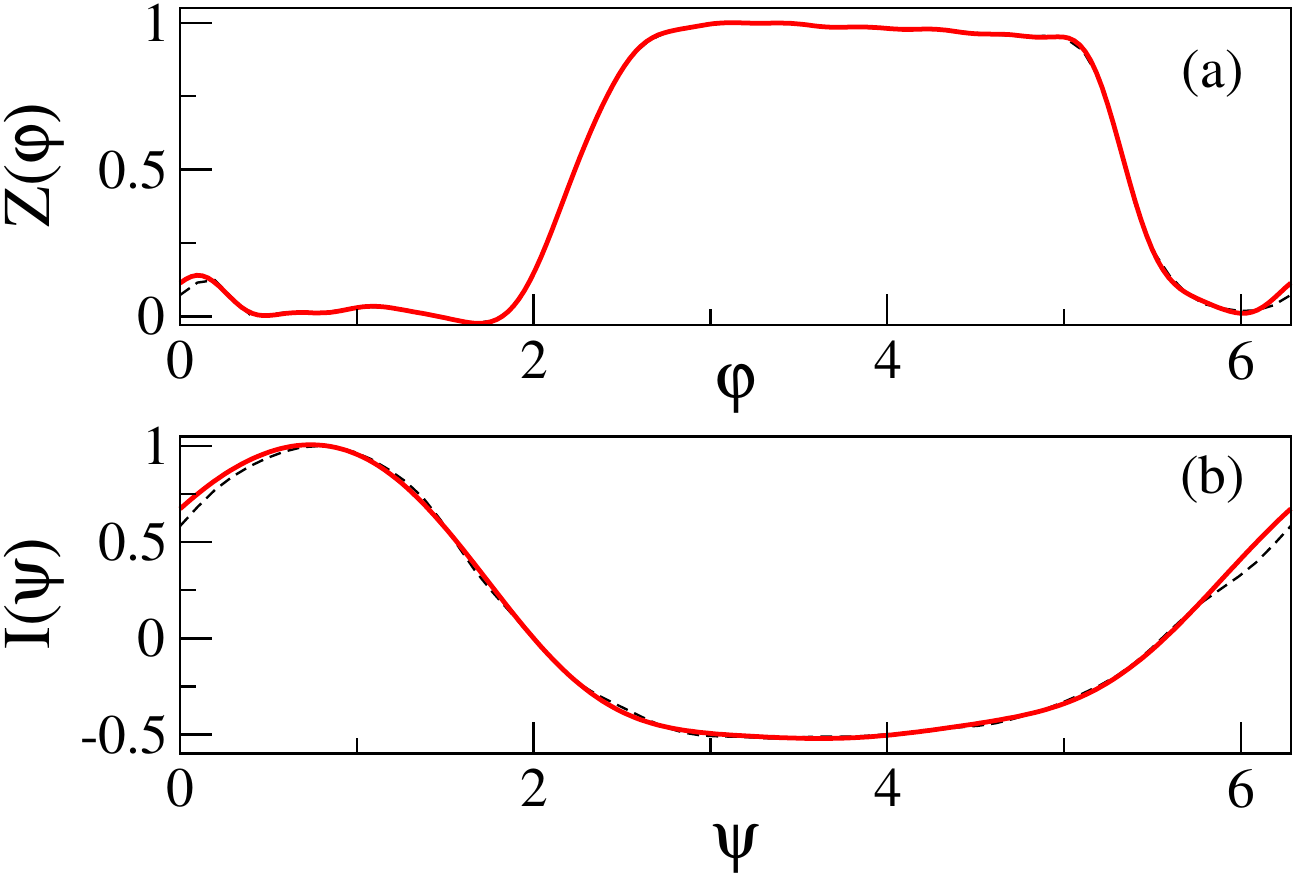}
\caption{Model phase response curve of the cardiac oscillator (a) and model respiratory 
force (b) are shown by bold red lines. Dashed lines in both panels indicate 
the corresponding curves obtained in experiments, cf.~\cite{Kralemann_et_al-13}.
}
\label{fig:ZandI}
\end{figure}

First we verify our approach using artificially generated data with known properties.
For this goal we use a simple phase model (\ref{Qcard}), where the coupling function 
$Q(\vp,\psi)$ is written in the Winfree form, i.e. as a product of the phase sensitivity 
function, or phase response curve (PRC), $Z(\vp)$, and forcing function $I(\psi)$.
Thus, introducing explicitly the coupling strength parameter $\e$, we write
\begin{equation}
\dot\vp=\w+\e Z(\vp)I(\psi)+\xi(t)\;.
\label{eq:mod1}
\end{equation}
Functions $Z(\vp)$, $I(\psi)$ are modeled by Fourier series of order 15 and 4, 
respectively, see Fig.~\ref{fig:ZandI},
in such a way that they resemble experimentally obtained curves, 
cf.~\cite{Kralemann_et_al-13}.
Instantaneous frequency of respiration was modeled as $\dot\psi=\w_r+\mu\nu$, where 
$\nu$ is an Ornstein-Uhlenbeck process, $\dot\nu=-\gamma_r \nu +\eta_1$. The random term
$\xi$ is given by the weighted sum of two components, i.e. of a low-pass and of
a band-pass filtered noise: $\xi=\lambda_1 \zeta_1+\lambda_2\zeta_2$, where 
$\dot\zeta_1=-\gamma\zeta_1+\eta_2$ and 
$\ddot \zeta_2+ \alpha\dot\zeta_2+\w_{bp}^2\zeta_2=\eta_3$. Here $\eta_k$ are independent
Gaussian white noises with zero mean: 
$\langle \eta_k(t)\eta_j(t')\rangle=\delta_{kj}\delta(t-t')$.

Solving stochastic differential
Eq.~(\ref{eq:mod1}), we generate the artificial series of R-peaks. 
Without loss of generality, we say that these peaks occur when phase $\vp$ attains
a multiple of $2\pi$. Thus, we obtain a point process $t_k$ such that 
$\vp(t_k)=2\pi k$. Correspondingly, we introduce series of RR-intervals $T_k=t_{k+1}-t_k$.
Similarly, solving the deterministic part
 of Eq.~(\ref{eq:mod1}), i.e.
\begin{equation}
\dot\vp^{(R)}=\w+\e Z(\vp)I(\psi)\;,
\label{eq:mod2}
\end{equation}
we generate a series of respiratory-related R-peaks, $t_k^{(R)}$ and corresponding 
intervals $T_k^{(R)}$~\footnote{Notice that although Eq.~(\ref{eq:mod2}) 
represents a deterministic part of  Eq.~(\ref{eq:mod1}), it remains a stochastic equation
due to presence in the respiratory phase $\psi$ of an Ornstein-Uhlenbeck 
process component. }. 
Finally, the non-respiratory related R-peaks,  
$t_k^{(NR)}$ and the interbeat intervals $T_k^{(NR)}$ are obtained via solution of 
\begin{equation}
\dot\vp^{(NR)}=\w+\xi(t)\;.
\label{eq:mod3}
\end{equation}
Thus, the data used for the disentanglement are: the times of R-peaks, $t_k$, 
and the respiratory phase and the frequency,
$\psi(t)$ and $\dot\psi(t)$, and in particular 
$\psi(t_k)=\psi_k$, and $\dot\psi(t_k)=\dot\psi_k$. 
Notice that in this test two latter series 
are obtained from equations, while in fact respiratory phase and frequency 
should be estimated from data, 
what certainly will introduce an additional error.
The respiratory-related and the non-respiratory-related components obtained via 
dynamic disentanglement, 
shall be compared with $t_k^{(R)}$ and $t_k^{(NR)}$, respectively.

\begin{figure}[!h]
\centering\includegraphics[width=3in]{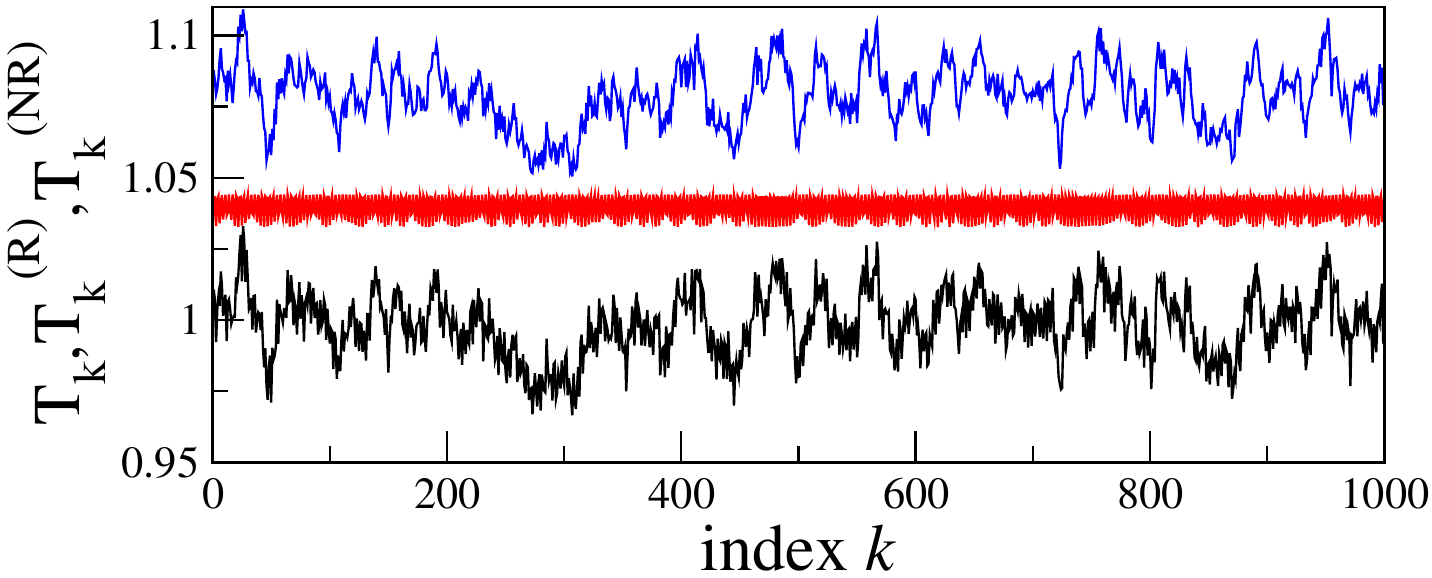}
\caption{A short epoch of the model data. 
From bottom to top: artificial sequence of interbeat interval, 
its respiratory-related, and non-respiratory related components. 
The latter two curves are shifted upwards for visibility.}
\label{fig:mod_tacho}
\end{figure}

\begin{figure}[!h]
\centering\includegraphics[width=3in]{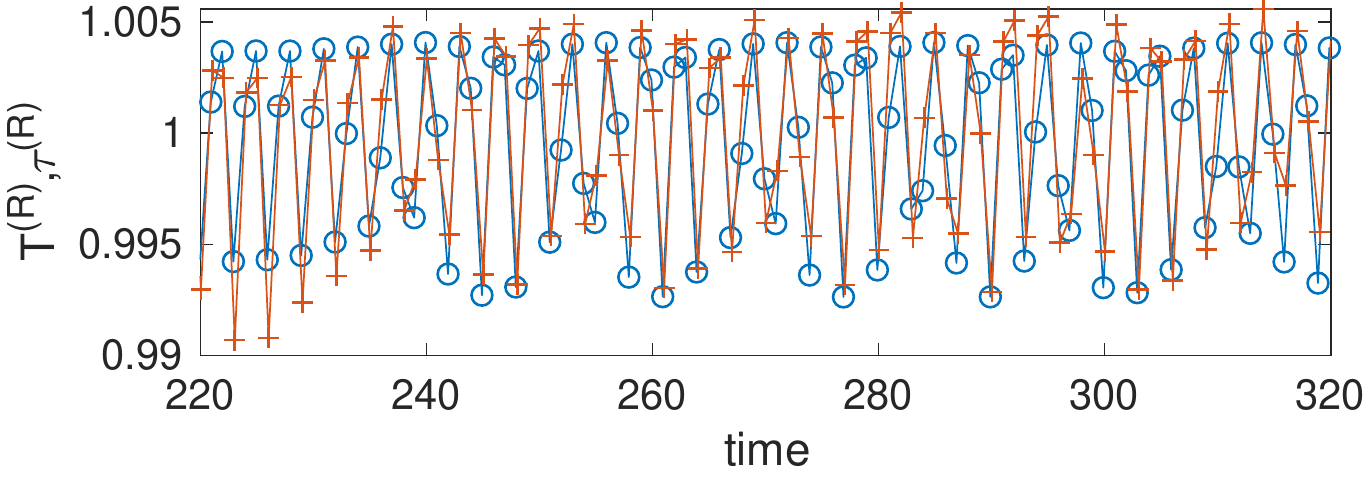}
\caption{True (blue circles) and recovered (red crosses) respiratory components of the HRV.
The first one is generated by the model, while the second one is obtained from the point 
process by means of constructing the coupling map (\ref{eq:fit}). 
}
\label{fig:mod_cleansed}
\end{figure}

\begin{figure}[!h]

\centering\includegraphics[width=0.48\textwidth]{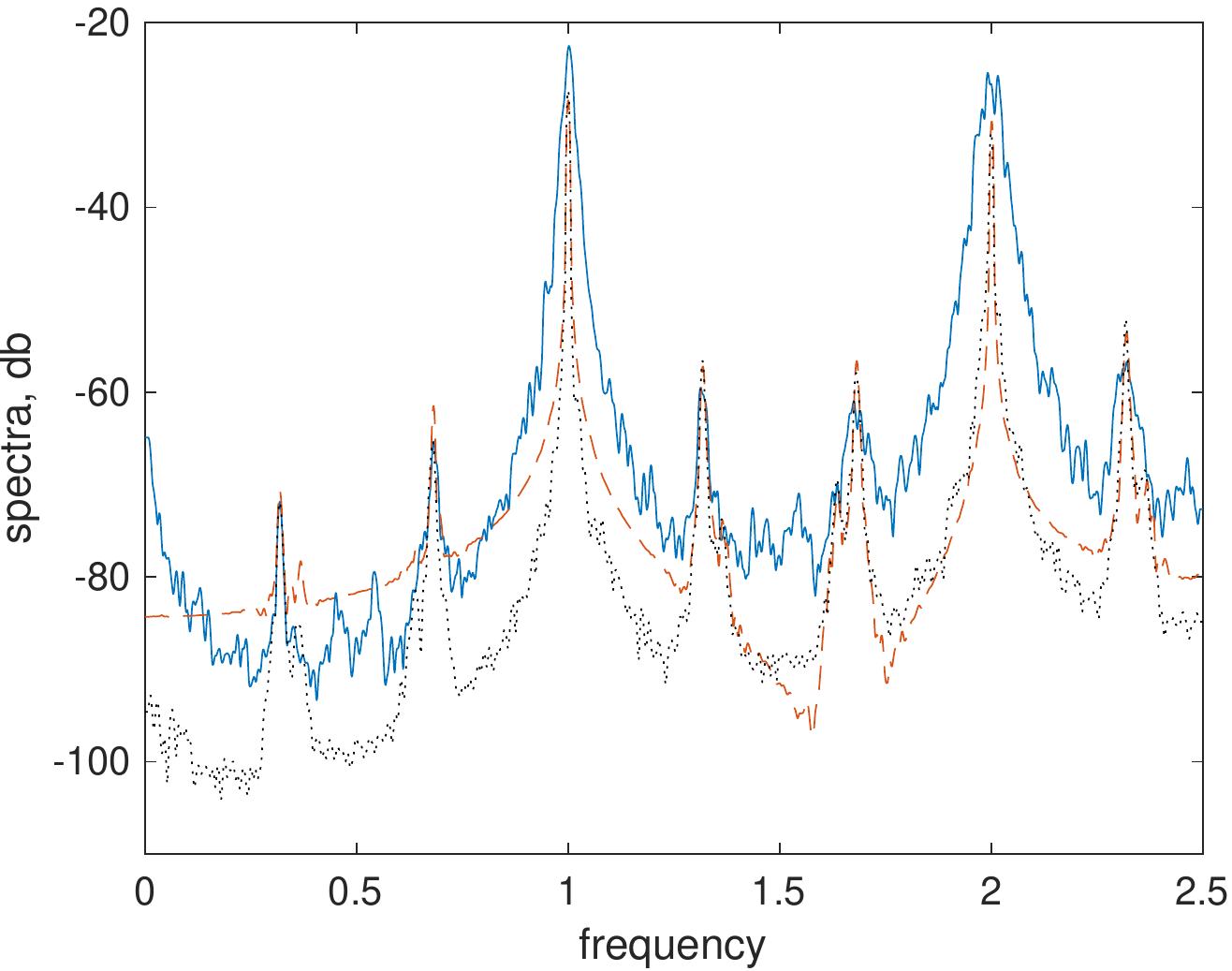}\hfill
\includegraphics[width=0.48\textwidth]{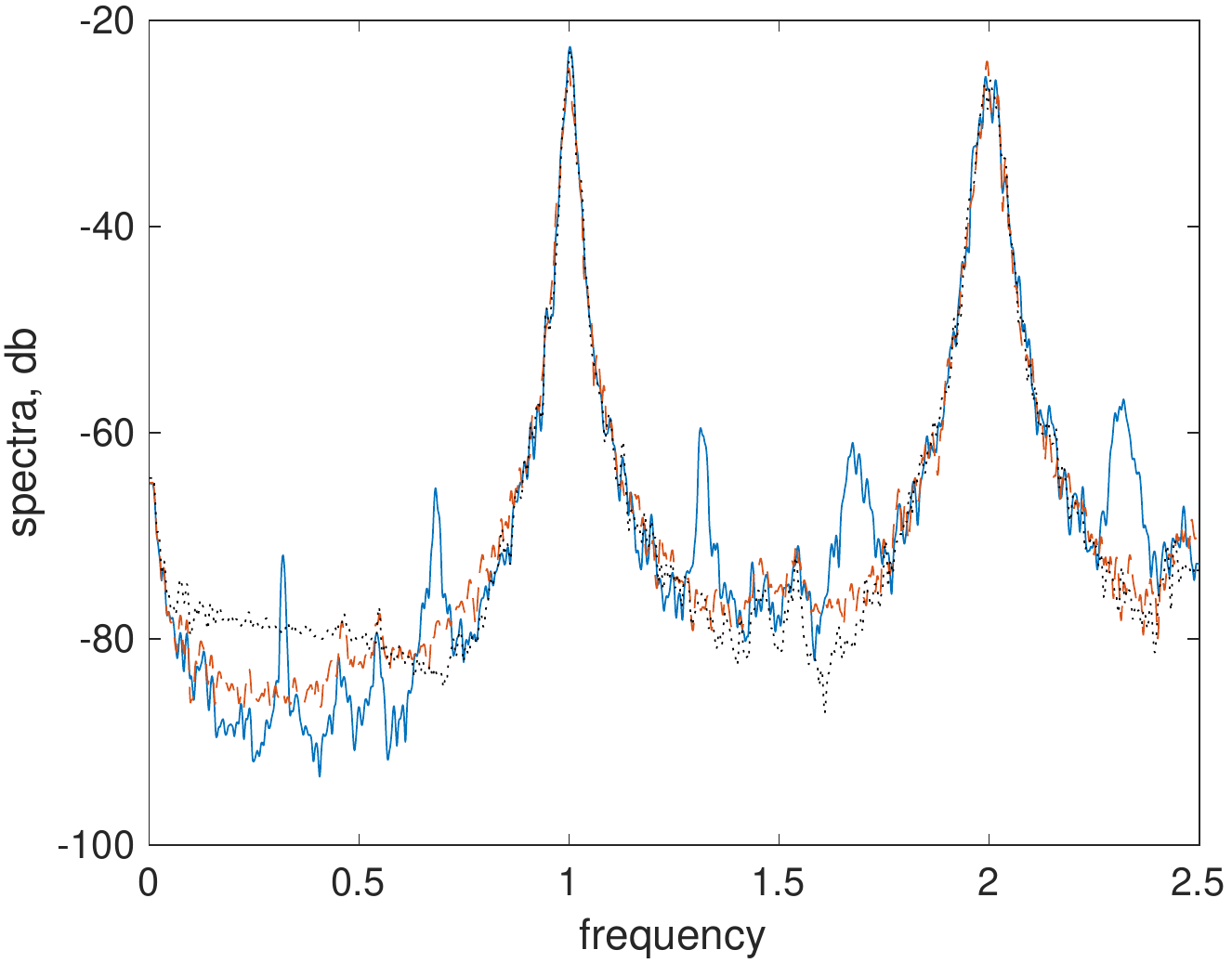}
\caption{Illustration of the disentanglement in the frequency domain.
Here we show the spectra of the point processes for the respiratory-related
component (left panel) and for the NR component (right panel). Blue solid line shows
the spectrum of model-generated series of instances of R-peaks $t_k$;
red dashed curves show the spectrum of the model-generated  
series $t_k^{(R)}$, $t_k^{(NR)}$, while 
the black dotted curves present the spectra of the R- and NR-components 
obtained from $t_k$ via disentanglement.}
\label{fig:mod_specR}
\end{figure}

Here we illustrate the model data and the disentanglement results for the following 
values of the parameters: $\w=2\pi$, $\w_r=2$, $\e=0.1$, $\w_{bp}=1.08\pi$, 
$\alpha=0.1$, $\gamma_r=0.1$, $\mu=0.02$, $\lambda_1=0.03$, $\lambda_2=0.02$,. 
The records used for the subsequent analysis contained about $10^4$ interbeat 
intervals, which correspond to about $2.5$ hours of natural heart beat.
The model data are illustrated in Fig.~\ref{fig:mod_tacho}. Here we show a short epoch
of the artificially generated sequences of RR-intervals. 
Figure~\ref{fig:mod_cleansed} presents the respiratory-related component, extracted 
with the help of our algorithm with $N_F=8$, $N_T=2$, compared to the true one, 
i.e. generated by the model.
Figure~\ref{fig:mod_specR} illustrate the results of the 
disentanglement in the frequency domain. 
Namely, here we present spectra of point processes  (Bartlett measure)~\cite{Bartlett-63}.
As expected, spectral peaks induced by respiration are enhanced in the R-component and 
suppressed in the NR-component.

We conclude the presentation of the technique by discussing a characterization 
of the quality of disentanglement. First we notice that, as it follows from 
Eqs.~(\ref{eq:mod1},\ref{eq:mod2},\ref{eq:mod3}) and as is expected for a
disentanglement of
independent components,
$\text{Var}(\dot\vp)=\text{Var}(\dot\vp^{(R)})+\text{Var}(\dot\vp^{(NR)})$, where 
the variance is defined as  $\text{Var}(x(t))=\langle (x-\langle x\rangle)^2\rangle$, 
$\langle (\cdot)\rangle=T_\Sigma^{-1}\int_0^{T_\Sigma} (\cdot) \dd t$, and $T_\Sigma$ is the time 
interval over which the averaging is performed. 
We expect that a similar relation for 
variances obtained from the interval series $T_k$, $T_k^{(R)}$, $T_k^{(NR)}$
shall be also valid, at least approximately. To compute the variance of the phase derivative
for a point process, we consider the phase linearly growing between the events, so that  
$\dot\vp(t)=2\pi/T_k$ for $t_k\le t\le t_{k+1}$, 
$k=1,\ldots,N$.
Then, for the variance we obtain 
\begin{equation}
\sigma^2=\frac{4\pi^2}{T_\Sigma}\sum_{k=1}^{N}\left(\frac{1}{T_k}-\frac{N}{T_\Sigma}\right)^2T_k\;,
\label{eq:ppvar}
\end{equation}  
where $T_\Sigma=\sum_kT_k$, and similarly for the respiratory-related and the 
non-respiratory-related 
components. We checked, for different $N_F,N_T$, that indeed
$(\sigma^2_R+\sigma^2_{NR})/\sigma^2 \approx 1$ (for $N_T\le 3$ the worst case was $0.97$).

\section{An application to human cardio-respiratory data}
Now we apply our algorithm to real data.  
For this goal we analyzed 26 multivariate records of ECG and respiration, 
registered in 17 healthy 
adults in supine position at rest, see \cite{Kralemann_et_al-13,Gallasch_et_al-96,Gallasch_et_al-97} 
for a detailed description of the subjects, experimental protocol, and measurement 
equipment\footnote{This study was performed with a high-grade equipment especially developed for RR 
variability measurements at sampling rate 1000 Hz and resolution 16 bit, with shielded ECG cables.
Notice that HRV measurements in medicine often do not meet such standards.
Low data sampling rates ($<1000$ Hz) and digital resolution ($<12$ bit) of commercial ECG 
equipment,  built not
for precise RR interval sampling but rather for low frequency ECG shape evaluation,
introduce artificial jitter and digitizing noise detrimental for precise variability determination.
}.
Since continuous phases $\vp(t),\psi(t)$ obtained in \cite{Kralemann_et_al-13} are 
available, we compare the approximate 
disentanglement performed with the help of Eqs.~(\ref{eq:cmap},\ref{eq:fit}) with 
the results obtained for continuous phases.
 
\begin{figure}[!h]
\centering\includegraphics[width=2.5in]{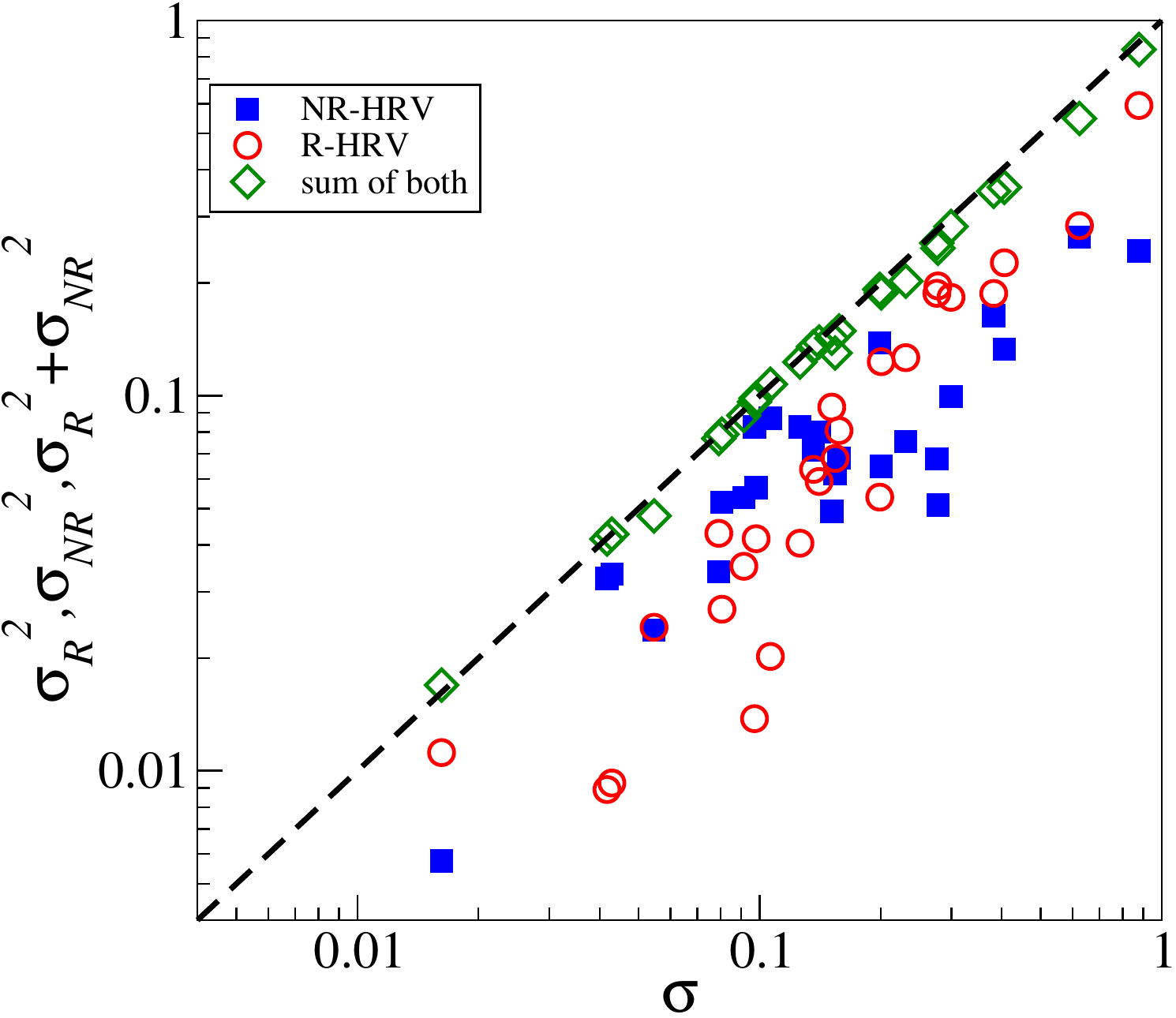}
\caption{Quality of the point-process based disentanglement for cardio-respiratory data from 
healthy subjects. Here we plot, for each experimental record, variances of the respiratory 
and non-respiratory-related components 
and their sum vs. variance of the original sequence of R-peaks. 
As expected for uncorrelated components,
the sum of variances of disentangled processes is very close to the 
variance of original data.}
\label{fig:realdata1}
\end{figure}

In order to quantify the quality of the disentanglement we compute 
$(\sigma^2_R+\sigma^2_{NR})/\sigma^2$ for all subjects with the help of 
Eq.~(\ref{eq:ppvar}). 
The results shown in Figs.~\ref{fig:realdata1} indicate that our algorithm works 
quite well.
Here we used $N_F=8$, $N_T=1$; for $N_T>1$ the quality of the disentanglement was
bad, probably because our point process series are quite short (about 400 heartbeats).  
\begin{figure}[!h]
\centering\includegraphics[width=2.5in]{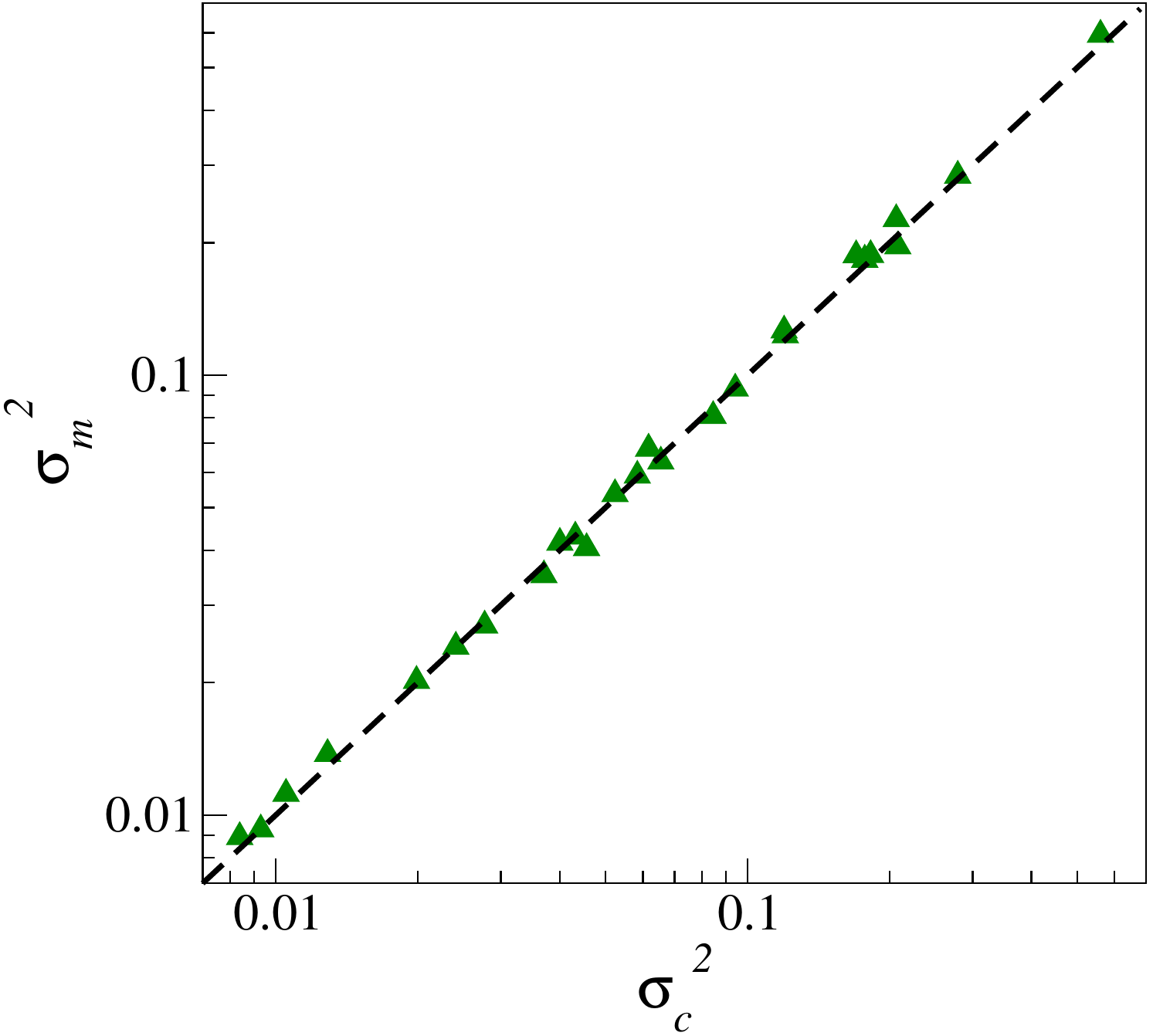}
\caption{The efficiency of the map-based disentanglement is confirmed by plotting the variance 
$\sigma_m^2$ of the map-cleansed respiratory-related component vs. the variance $\sigma_c^2$
of the continuously-cleansed respiratory-related component.
}
\label{fig:realdata2}
\end{figure}
Next, we compare the variance obtained from map-cleansed intervals with 
the variance for continuously-cleansed data, see Fig.~\ref{fig:realdata2}.
Both are in a good agreement. 

\begin{figure}[!h]
\centering\includegraphics[width=3.5in]{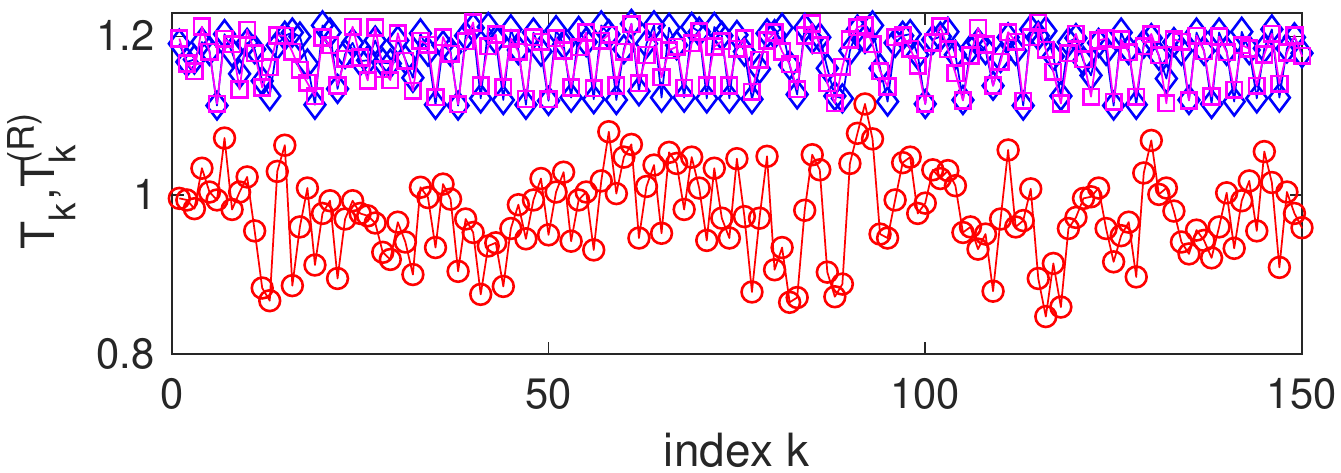}
\caption{Original series of RR-intervals, $T_k$ (in seconds), of a healthy subject (red circles)
and its respiratory-related component, $T_k^{(R)}$, obtained either via continuous
cleansing (blue diamonds) or by
map cleansing (magenta squares). Both cleansed time series are shifted
upwards for visibility.}
\label{fig:realdata3}
\end{figure}

An example of disentanglement (for a particular recording, data set 3) 
is shown in Fig.~\ref{fig:realdata3}.
Here we show the original series of RR-intervals and the two cleansed data sets, one obtained
via the disentanglement with the continuous phase and one obtained using only the R-peaks.
For better visibility, two cleansed data sets are shifted upwards by $0.2$s, their overlap
indicates that the discrete dynamical disentanglement works well.

\section{Conclusion}
To summarize, we have presented a general approach that allows us, 
by means of the reconstruction of the dynamics of
a driven oscillator, to predict its ``virtual dynamics'' in
which some of its inputs are cut off. 
The described disentanglement procedure 
differs from known mode decomposition algorithms, because it operates 
not with the given time series, but
with a reconstructed phase dynamics equation. In this paper we 
focused on the extension of the dynamical disentanglement
 to the case when 
the output of the investigated oscillator is a sequence of events (a point process),
 so that its 
instantaneous phase 
can be hardly estimated, while the observed input is relatively slow and smooth 
process, suitable
for the phase estimation. We applied this approach to analyze human heart rate 
variability, where the available time series are a respiration signal and heart beat
events. More precisely,
we disentangled the respiratory-related variability, known to be mediated by the 
vagus nerve only,  from that due to other sources. 
Using both model data as well as instantaneous phases derived from an 
electrocardiogram, we have shown that 
our approximate procedure yields quite good results. 

The developed technique can find applications in physiological as well as clinical studies.  
Indeed, quantification of different components of HRV is already
an important diagnostic and prognostic tool in cardiology \cite{Moser1994,Malik_et_al-96}.
Several circulatory diseases show a strong difference in prognosis depending on vagal activity.
As already mentioned, the simple spectral methods applied to RR interval analysis in many 
clinical studies is not performing well in separating vagal respiratory and other components 
of HRV. Since our technique provides respiratory-related variability cleansed from the effect 
of noise and other, unobserved rhythms, quantification of RSA and hence vagal tone from 
disentangled data is more precise.
The separation of respiratory  and non-respiratory components is physiologically and 
clinically especially important in slow breathing ($<0.12$ Hz), where the vagal RSA
intermixes with slower rhythms like blood pressure rhythm, which derive 
from sympathetic and vagal components. Under such conditions the two
components of autonomic nervous system activity cannot be separated with 
linear models \cite{Cysarz_et_al-04}.

In Ref.~\cite{Topcu_et_al-18} we compared 
performance of different RSA measures applied to original and 
cleansed series. However, there the 
disentanglement was performed using instantaneous continuous phases. 
Now we show that the practical algorithm that operates not 
with a continuous ECG, but only with R-peaks,
provides nearly the same results. This finding opens a way to practical use.
We anticipate that the developed technique can be also used in neuroscience, e.g. for analysis 
of spiking of sensory neurons in response to a slowly varying stimulus.

As a subject of future research, we mention a case of more than one observed input.
Theoretically, it is not so difficult to perform the phase dynamics reconstruction 
for multivariate data; however, data requirements increase essential so that 
reconstruction of a network of more than 3 oscillators becomes unfeasible, 
see~\cite{Kralemann-Pikovsky-Rosenblum-14}. 
Another interesting extension would be the case
of non-oscillatory inputs, when parameterization of these inputs by a phase does not work. 
A possible solution for both problems might be reconstruction of the phase dynamics in the 
Winfree form, i.e. when the coupling function can be presented as a product 
of the phase response 
curve and of the driving signal, cf.~Eq.~(\ref{eq:mod1}).  
\vskip6pt

\enlargethispage{20pt}

MR, MM, and AP acknowledge financial support from the European Union's Horizon
2020 research and innovation programme under the Marie
Sklodowska-Curie Grant Agreement No. 642563 (COSMOS).
Development of methods presented in Section 4 was supported by
the Russian Science Foundation under Grant No. 17-12-01534.


\begin{thebibliography}{10}
\expandafter\ifx\csname urlstyle\endcsname\relax
  \providecommand{\doi}[1]{(doi:\discretionary{}{}{}#1)}\else
  \providecommand{\doi}{(doi:\discretionary{}{}{}\begingroup
  \urlstyle{rm}\Url)}\fi

\bibitem{Fukunaga-90}
Fukunaga K. 1990 \emph{Introduction to Statistical Pattern Recognition}.
\newblock Amsterdam: Elsevier.

\bibitem{Huang-98}
{Huang} NE, {Shen} Z, {Long} SR, {Wu} MC, {Shih} HH, {Zheng} Q, {Yen} NC,
  {Tung} CC, {Liu} HH. 1998 {The empirical mode decomposition and the Hilbert
  spectrum for nonlinear and non-stationary time series analysis}.
\newblock \emph{Proceedings of the Royal Society of London Series A}
  \textbf{454}, 903--998.

\bibitem{Jolliffe-2002}
Jolliffe I. 2002 \emph{Principal Component Analysis}.
\newblock Berlin: Springer.

\bibitem{Flandrin-04}
Flandrin P, Rilling
  as a filter bank.
\newblock \emph{IEEE Signal Processing Lett.} \textbf{11}, 112--114.

\bibitem{Feldman-11}
Feldman M. 2011 \emph{Hilbert Transform Applications in Mechanical Vibration}.
\newblock UK: Wiley.

\bibitem{PhysRevE.92.032916}
Iatsenko D, McClintock PVE, Stefanovska A. 2015 Nonlinear mode decomposition: A
  noise-robust, adaptive decomposition method.
\newblock \emph{Phys. Rev. E} \textbf{92}, 032916.

\bibitem{van_der_Pol_van_der_Mark-28}
van~der Pol B, van~der Mark. 1928 The heartbeat considered as a relaxation
  oscillation and an electrical model of the heart.
\newblock \emph{Phil. Mag.} \textbf{6}, 763--775.

\bibitem{Schaefer-Rosenblum-Kurths-Abel-98}
Sch\"afer C, Rosenblum MG, Kurths J, Abel HH. 1998 Heartbeat synchronized with
  ventilation.
\newblock \emph{Nature} \textbf{392}, 239--240.

\bibitem{Mrowka-Patzak-Rosenblum-00}
Mrowka R, Patzak A, Rosenblum MG. 2000 Qantitative analysis of
  cardiorespiratory synchronization in infants.
\newblock \emph{Int. J. of Bifurcation and Chaos} \textbf{10}, 2479--2488.

\bibitem{PhysRevLett.85.4831}
Stefanovska A, Haken H, McClintock PVE, Ho\ifmmode \check{z}\else
  \v{z}\fi{}i\ifmmode~\check{c}\else \v{c}\fi{} M,
  Bajrovi\ifmmode~\acute{c}\else \'{c}\fi{} F, Ribari\ifmmode~\check{c}\else
  \v{c}\fi{} S. 2000 Reversible transitions between synchronization states of
  the cardiorespiratory system.
\newblock \emph{Phys. Rev. Lett.} \textbf{85}, 4831--4834.

\bibitem{Rosenblum_et_al-02}
Rosenblum MG, Cimponeriu L, Bezerianos A, Patzak A, Mrowka R. 2002
  Identification of coupling direction: Application to cardiorespiratory
  interaction.
\newblock \emph{Phys. Rev. E} \textbf{65}, 041909.

\bibitem{Mrowka_et_al-03}
Mrowka R, Cimponeriu L, Patzak A, Rosenblum M. 2003 Directionality of coupling
  of physiological subsystems - age related changes of cardiorespiratory
  interaction during different sleep stages in babies.
\newblock \emph{American J. of Physiology Regul. Comp. Integr. Physiol.}
  \textbf{145}, R1395--R1401.

\bibitem{Kralemann_et_al-13}
Kralemann B, Fr\"uhwirth M, Pikovsky A, Rosenblum M, Kenner T, Schaefer J,
  Moser M. 2013 In vivo cardiac phase response curve elucidates human
  respiratory heart rate variability.
\newblock \emph{Nature Communications} \textbf{4}, 2418.

\bibitem{Iatsenko_et_al-13}
Iatsenko D, Bernjak A, Stankovski T, Shiogai Y, Jane OLP, Clarkson PBM,
  McClintock PVE, Stefanovska A. 2013 Evolution of cardiorespiratory
  interactions with age.
\newblock \emph{Philosophical Transactions of the Royal Society of London A}
  \textbf{371}.

\bibitem{Eckberg-83}
Eckberg DL. 1983 Human sinus arrhythmia as an index of vagal cardiac outflow.
\newblock \emph{J. Appl. Physiol.} \textbf{54}, 961--966.

\bibitem{Berntson-Cacioppo-Quigley-93}
Berntson GG, Cacioppo JT, Quigley KS. 1993 Respiratory sinus arrhythmia:
  {A}utonomic origins, physiological mechanisms, and psychological
  implications.
\newblock \emph{Psychophysiology} \textbf{30}, 183--196.

\bibitem{Eckberg-03}
Eckberg DL. 2003 The human respiratory gate.
\newblock \emph{The Journal of physiology} \textbf{548(Pt 2)}, 339--352.

\bibitem{Billman-11}
Billman G. 2011 Heart rate variability -- a historical perspective.
\newblock \emph{Front Physiol.} \textbf{2}, 86.

\bibitem{Beauchaine-15}
Beauchaine TP. 2015 Respiratory sinus arrhythmia: {A} transdiagnostic biomarker
  of emotion dysregulation and psychopathology.
\newblock \emph{Current opinion in psychology} \textbf{3}, 43--47.

\bibitem{Lehofer1999}
Lehofer M, Moser M, Hoehn-Saric R, McLeod D, Hildebrandt G, Egner S,
  Steinbrenner B, Liebmann P, Zapotoczky HG. 1999 {Influence of age on the
  parasympatholytic property of tricyclic antidepressants}.
\newblock \emph{Psychiatry Research} \textbf{85}, 199--207.

\bibitem{Moser1998}
Moser M, Lehofer M, Hoehn-Saric R, McLeod DR, Hildebrandt G, Steinbrenner B,
  Voica M, Liebmann P, Zapotoczky HG. 1998 {Increased heart rate in depressed
  subjects in spite of unchanged autonomic balance?}
\newblock \emph{Journal of Affective Disorders} \textbf{48}, 115--124.

\bibitem{Tracey2002}
Tracey KJ. 2002 {The inflammatory reflex}.
\newblock \emph{Nature} \textbf{420}, 853--9.

\bibitem{Olofsson_et_al-12}
Olofsson P, Rosas-Ballina M, Levine Y, Tracey K. 2012 Rethinking inflammation:
  neural circuits in the regulation of immunity.
\newblock \emph{Immunol Rev} \textbf{248}, 188--204.

\bibitem{Andersson-Tracey-12}
Andersson U, Tracey K. 2012 Neural reflexes in inflammation and immunity.
\newblock \emph{J. Exp. Med.} \textbf{209}, 1057--1068.

\bibitem{Rosas-Ballina-Tracey-09}
Rosas-Ballina M, Tracey K. 2009 Cholinergic control of inflammation.
\newblock \emph{J. Intern. Med.} \textbf{265}, 663--679.

\bibitem{Nathan-Ding-2010}
Nathan C, Ding A. 2010 Nonresolving inflammation.
\newblock \emph{Cell} \textbf{140}, 871--882.

\bibitem{Donchin_et_al-92}
Donchin Y, Constantini S, Szold A, Byrne EA, Porges SW. 1992 Cardiac vagal tone
  predicts outcome in neurosurgical patients.
\newblock \emph{Crit. Care Med.} \textbf{20}, 942.

\bibitem{Moser2006}
Moser M, Fr\"uhwirth M, Penter R, Winker R. 2006 Why life oscillates -- from
  topographical towards a functional chronobiology.
\newblock \emph{Cancer Cause Control} \textbf{17}, 591--599.

\bibitem{Das2011}
Das UN. 2011 Can vagus nerve stimulation halt or ameliorate rheumatoid
  arthritis and lupus?
\newblock \emph{Lipids in Health and Disease} \textbf{10}, 19.

\bibitem{Chow_et_al-14}
Chow E, Iqbal A, Bernjak A, Ajjan R, Heller SR. 2014 Effect of hypoglycaemia on
  thrombosis and inflammation in patients with type 2 diabetes.
\newblock \emph{Lancet} \textbf{383}, S35.

\bibitem{Topcu_et_al-18}
\c{C} Top\c{c}u, Fr\"uhwirth M, Moser M, Rosenblum M, Pikovsky A. 2018
  Disentangling respiratory sinus arrhythmia in heart rate variability records.
\newblock \emph{Physiological Measurements} \textbf{39}, 054002.

\bibitem{Winfree-67}
Winfree AT. 1967 Biological rhythms and the behavior of populations of coupled
  oscillators.
\newblock \emph{J. Theor. Biol.} \textbf{16}, 15.

\bibitem{Kuramoto-84}
Kuramoto Y. 1984 \emph{Chemical Oscillations, Waves and Turbulence}.
\newblock Berlin: Springer.

\bibitem{Pikovsky-Rosenblum-Kurths-01}
Pikovsky A, Rosenblum M, Kurths J. 2001 \emph{Synchronization. A Universal
  Concept in Nonlinear Sciences.}
\newblock Cambridge: Cambridge University Press.

\bibitem{Kralemann-Pikovsky-Rosenblum-11}
Kralemann B, Pikovsky A, Rosenblum M. 2011 Reconstructing phase dynamics of
  oscillator networks.
\newblock \emph{Chaos} \textbf{21}, 025104.

\bibitem{Kralemann-Pikovsky-Rosenblum-14}
Kralemann B, Pikovsky A, Rosenblum M. 2014 Reconstructing effective phase
  connectivity of oscillator networks from observations.
\newblock \emph{New Journal of Physics} \textbf{16}, 085013.

\bibitem{Rosenblum-Pikovsky-18}
Rosenblum M, Pikovsky A. 2018 Efficient determination of synchronization
  domains from observations of asynchronous dynamics.
\newblock \emph{Chaos} \textbf{28}, 106301.

\bibitem{Note1}
For integration we used the Euler scheme; for initial conditions both $\varphi
  ^{(R)}$ and $\varphi ^{(NR)}$ we set to zero at the instant of the first
  R-peak in the original data set. Since the coupling function $Q$ is given on
  a grid, spline interpolation was used to compute $Q(\varphi ,\psi )$ for
  arbitrary $\varphi ,\psi $.

\bibitem{Widjaja_et_al-14}
Widjaja D, Caicedo A, Vlemincx E, Van~Diest I, Van~Huffel S. 2014 Separation of
  respiratory influences from the tachogram: A methodological evaluation.
\newblock \emph{PLOS ONE} \textbf{9}, 1--11.

\bibitem{Kuo-Kuo-16}
Kuo J, Kuo CD. 2016 Decomposition of heart rate variability spectrum into a
  power-law function and a residual spectrum.
\newblock \emph{Front. Cardiovasc. Med.} \textbf{3}, 16.

\bibitem{Note2}
For a high-resolution measurement phase and frequency of respiration are given
  as a time series with a small time step. Therefore, their values at
  $t_2^{(R)}$ can be obtained, e.g. by linear interpolation between two closest
  data points.

\bibitem{Note3}
Notice that although Eq.~(\ref {eq:mod2}) represents a deterministic part of
  Eq.~(\ref {eq:mod1}), it remains a stochastic equation due to presence in the
  respiratory phase $\psi $ of an Ornstein-Uhlenbeck process component.

\bibitem{Bartlett-63}
Bartlett M. 1963 The spectral analysis of point processes.
\newblock \emph{J. R. Statist. Soc. Ser. B} \textbf{29}, 264--296.

\bibitem{Gallasch_et_al-96}
Gallasch E, Rafolt D, Moser M, Hindinger J, Eder H, Wiesspeiner G, Kenner T.
  1996 Instrumentation for assessment of tremor, skin vibrations, and
  cardiovascular variables in {MIR} space missions.
\newblock \emph{IEEE Trans. Biomed. Eng.} \textbf{43}, 328--333.

\bibitem{Gallasch_et_al-97}
Gallasch E, Moser M, Kozlovskaya I, Kenner T, Noordergraaf A. 1997 Effects of
  an eight-day space flight on microvibration and physiological tremor.
\newblock \emph{Amer. J Physiol, Regul Integr Card} \textbf{273}, R86--R92.

\bibitem{Note4}
This study was performed with a high-grade equipment especially developed for
  RR variability measurements at sampling rate 1000 Hz and resolution 16 bit,
  with shielded ECG cables. Notice that HRV measurements in medicine often do
  not meet such standards. Low data sampling rates ($<1000$ Hz) and digital
  resolution ($<12$ bit) of commercial ECG equipment, built not for precise RR
  interval sampling but rather for low frequency ECG shape evaluation,
  introduce artificial jitter and digitizing noise detrimental for precise
  variability determination.

\bibitem{Moser1994}
Moser M, Lehofer M, Sedminek A, Lux M, Zapotoczky HG, Kenner T, Noordergraaf A.
  1994 Heart rate variability as a prognostic tool in cardiology. a
  contribution to the problem from a theoretical point of view.
\newblock \emph{Circulation} \textbf{90}, 1078--1082.

\bibitem{Malik_et_al-96}
Malik M, Bigger JT, Camm J, Kleiger RE, Malliani A, Moss AJ, Schwartz PJ. 1996
  Heart rate variability: standards of measurement, physiological
  interpretation, and clinical use.
\newblock \emph{Eur. Heart J.} \textbf{17}, 354--381.

\bibitem{Cysarz_et_al-04}
Cysarz D, von Bonin D, Lackner H, Heusser P, Moser M, Bettermann H. 2004
  Oscillations of heart rate and respiration synchronize during poetry
  recitation.
\newblock \emph{Amer J Physiol - Heart Circ Phys} \textbf{287}, H579--H587.

\end{thebibliography}
%

\end{document}